\def\year{2020}\relax
\documentclass[letterpaper]{article} % DO NOT CHANGE THIS
\usepackage{aaai20}  % DO NOT CHANGE THIS
\usepackage{times}  % DO NOT CHANGE THIS
\usepackage{helvet} % DO NOT CHANGE THIS
\usepackage{courier}  % DO NOT CHANGE THIS
\usepackage[hyphens]{url}  % DO NOT CHANGE THIS
\usepackage{graphicx} % DO NOT CHANGE THIS
\usepackage{color,soul}
\usepackage{amsmath}
\usepackage{placeins}
\usepackage{makecell}
\usepackage{graphicx}  % DO NOT CHANGE THIS
\title{Physics guided deep learning generative models for crystal materials discovery}
\author{Yong Zhao, Edirisuriya MD Siriwardane,  \Large \textbf{Jianjun Hu}\textsuperscript{\rm 1}\thanks{Corresponding author}\\ % All authors must be in the same font size and format. Use \Large and \textbf to achieve this result when breaking a line
\textsuperscript{\rm 1}Department of Computer Science and Engineering\\
University of South Carolina\\
550 Assembly Street\\
Columbia, SC, 29201\\
jianjunh@cse.sc.edu % email address must be in roman text type, not monospace or sans serif
}
 
\begin{document}

\maketitle

\begin{abstract}

Deep learning based generative models such as deepfake have been able to generate amazing images and videos. However, these models may need significant transformation when applied to generate crystal materials structures in which the building blocks, the physical atoms are very different from the pixels. Naively transferred generative models tend to generate a large portion of physically infeasible crystal structures that are not stable or synthesizable. Herein we show that by exploiting and adding physically oriented data augmentation, loss function terms, and post processing, our deep adversarial network (GAN) based generative models can now generate crystal structures with higher physical feasibility and expand our previous models which can only create cubic structures. 

\end{abstract}

\section{Introduction}

\noindent Discovering novel materials for high-temperature superconductors, high-capacity batteries and solar panels are highly desirable in a wide variety of industries. Traditionally, this new materials exploration step is mostly done via a trial-and-error tinkering process, which is both tedious and costly for both experimental exploration or computational exploration using e.g. first principle calculations. 

With the emergence of high performance computing and the machine learning and deep learning techniques, there are two promising approaches for discovering new materials: (1) generating new chemical formula/compositions \cite{dan_generative_2020} and then use computational crystal structure prediction algorithms \cite{oganov2011evolutionary,oganov2019structure,hu2021alphacrystal} to predict their structures and estimate their stability and synthesizability before final experimental validation \cite{song_computational_2021}; (2) structure-oriented generative machine learning models such as our CubicGAN \cite{zhao2021high} and other autoencoder and generative adversarial network based models \cite{ren2020inverse,noh2019inverse} as surveyed in \cite{noh2020machine}. In our recent work \cite{zhao2021high}, we have demonstrated that our deep neural network based generative adversarial network (GAN) can be trained to generate novel stable crystal cubic materials with novel prototypes which cannot be easily obtained using simple element substitution. More than 35,000 generated structures are accessible at our Carolina Materials database~\cite{cmd2021}. We also show that our model is able to recover almost all known cubic materials. While promising, we also find that there are several limitations for this study. First, the deep generative models can only work for cubic structures where the fractional coordinates are all multiples of 0.25. The complexity to generate crystal structures with arbitrary fractional coordinates are non-trivial. Second, our models are only able to generate structures of cubic crystal systems while there are many non-cubic structures that are out of our scope. Finally, we find that the success rate is still not satisfactory as our models still tend to generate a large portion of non-physical or unstable hypothetical structures. Considering that atoms are very different from pixels as used in the development of most GAN models \cite{westerlund2019emergence}, we realize that it is critical to incorporate more physical rules or constraints into our deep generative models to enable more effective generation of physically meaningful hypothetical structures. 

Our contribution includes the following:

\begin{itemize}
\item We develop deep generative models for crystal structure generation for non-cubic materials compared to our previous work
\item We propose a novel mechanism for data augmentation by exploiting an intrinsic physical property of the crystal structures, the symmetry operations
\item We develop a physics guided loss function to push the generative models to create more physically meaningful structures.
\item We develop a physics-guided post processing step to fine-tune our generated materials structures. 

\end{itemize}

\section{Methods}

\subsection{Physics guided generative models}

\subsubsection{(a) Framework of the GAN model for crystal generation}

\begin{figure*}[htb!]
	\centering
	\includegraphics[width=0.85\linewidth]{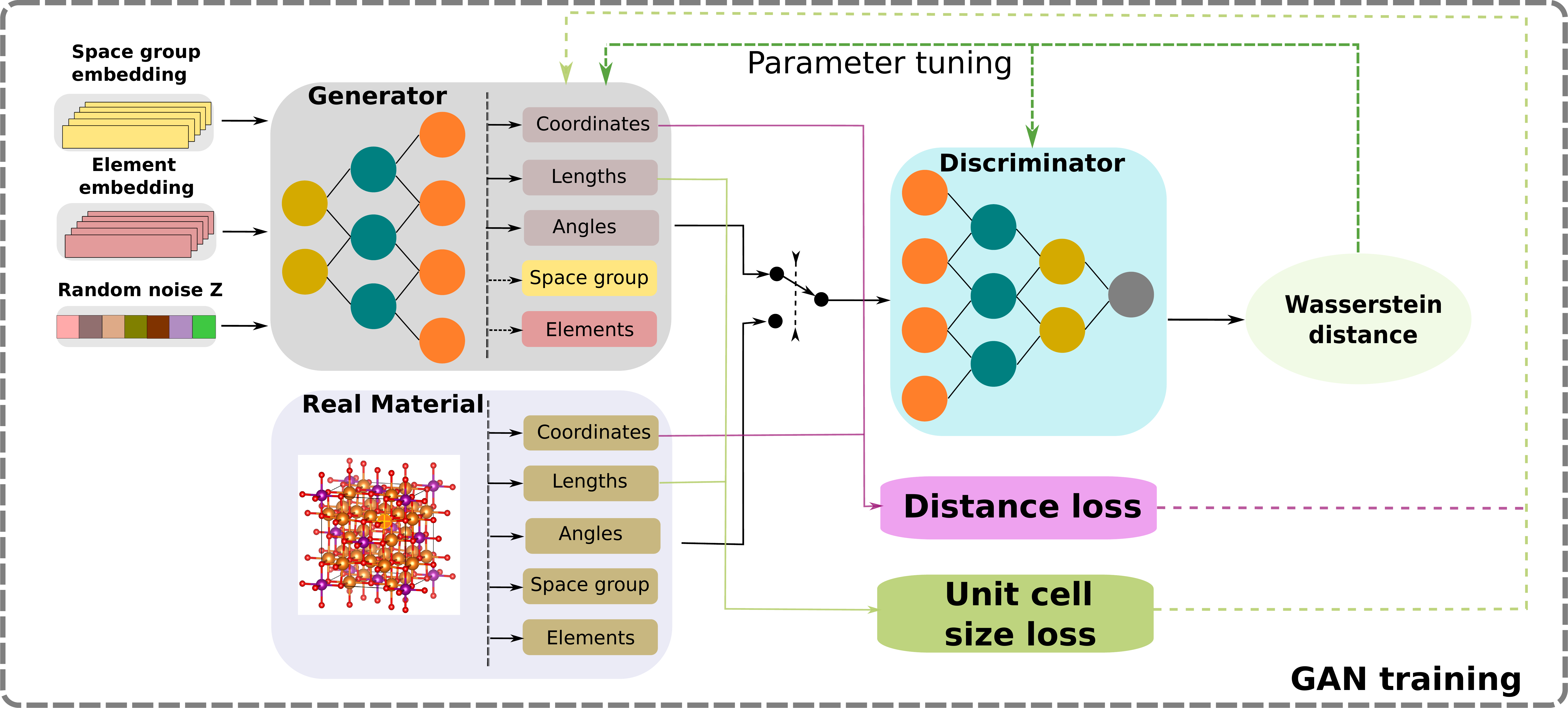}	
  	\caption {The diagram of physics guided GAN model for crystal structure generation.}
  	\label{fig:framework}
\end{figure*}

Our physics guided deep generative model for crystal structure generation is based on the Wasserstein generative adversarial network framework \cite{arjovsky2017wasserstein} as shown in Figure\ref{fig:framework}. It is composed of a generator, a discriminator and physics guided data augmentation mechanism, a physics guided loss terms added to the Wasserstein distance, and a physics informed post-processing step to cluster groups atoms into a single atom. The network configuration can be found in \cite{zhao2021high} as used in our CubicGAN model. Here we focus on the newly added physics guided components of our generative model.

\FloatBarrier

\subsubsection{(b) Self-augmentation of base atom sites}

\begin{figure}[htb!]
	\centering
	\includegraphics[width=\linewidth]{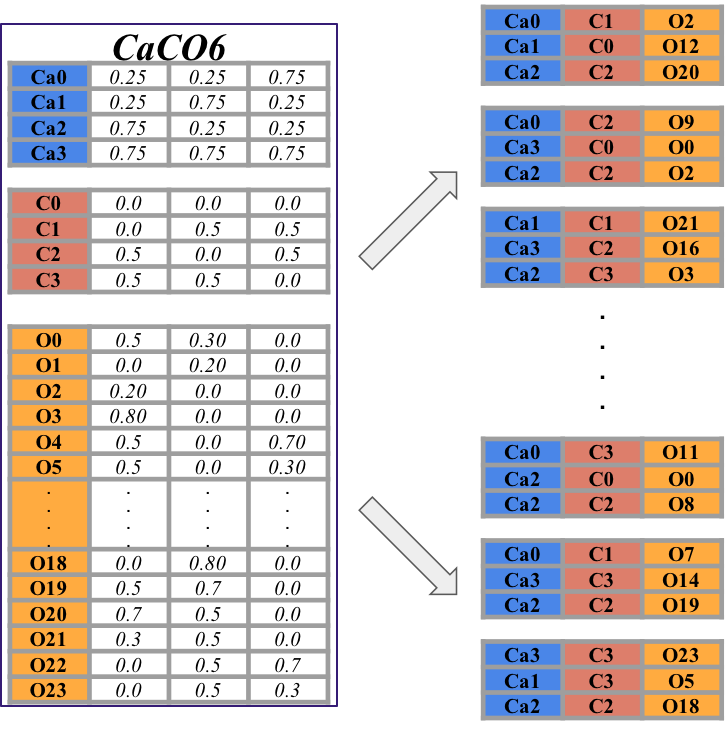}	
  	\caption {The diagram of self-augmentation mechanism. Ca, C and O are three elements in material of CaCO6, which have 4, 4, 24 atom positions, respectively. Each time, we randomly pick one of atom position for each element to form a set of base atom sites. Each set of base sites can be transformed into the same full set of 32 atom sites in unit cell.}
  	\label{fig:self-aug}
\end{figure}

In CubicGAN~\cite{zhao2021high}, we select a set of base (non-equivalent) atom sites as part of their inputs to the neural networks and the training datasets contain materials of which all the atom fractional coordinates have a multiplicative factor of 0.25. With a large number of materials, the generator can learn how to assign different fractional coordinates with a multiplicative factor of 0.25 combining with other crystal information, such as lattice and element properties. However, we only focus on three space groups (Fm$\bar{3}$m, F$\bar{4}$3m and Pm$\bar{3}$m) in that study. Compared to the 230 space groups in crystal materials, it is critical to develop generative models for other space groups as well for generating materials. However, most other space groups only have relatively small number of materials and we need a way to increase the number of materials for each space group. As discovered in our previous work \cite{zhao2021high}, the most challenging part of crystal generative models is learning the atom positions. Symmetry operations of space groups can generate dozens of unique atom positions for a single base atom position, which causes a large number of atoms in a single unit cell for high-symmetry space groups when the generated coordinates have slight deviation from the true symmetry sites. The larger the number of atoms in the unit cell, the higher the difficulty to optimize its structure successfully. Here, we propose a novel self-augmentation method to increase the training samples for space groups with insufficent samples by exploiting the physical spatial symmetry of the crystal structures. We take CaCO6 (F$\bar{4}$3m) as an example from our Carolina Materials Database~\cite{cmd2021} to demonstrate the self-augmentation mechanism as shown in Figure~\ref{fig:self-aug}.

The material CaCO6 has 32 atoms in the unit cell, of which both Carbon and Calcium elements have 4 unique atom positions and Oxygen has 24 unique ones. For each element, we can choose anyone as the base atom site. The self-augmentation process is to randomly select one base atom site for different elements to create a sample. The process can be repeated multiple times. In this work, we use 3 sets of base atom positions representations for one sample and each materials can generate multiple number of samples. We use 32 in our model. Hence, one materials can be augmented 32 times with different coordinate combinations with the same lattice, space group encoding and element properties. 

\subsubsection{(c) Physics guided losses}
Different from our previous model\cite{zhao2021high}, here we aim to generate crystal structures without constraining the values of the fractional coordinates of Wyckoff sites. However, this choice makes it easy to generate structures with atoms concentrated together or colliding each other due to the symmetry operations, which violates the basic physical rules of atomic interactions. To address this issue, we design a distance based loss to encourage atoms to keep away from each other as shown in below equation:
\begin{equation*}
    \begin{split}
        L_{dist} &=\lambda_{inter}\times L_{inter} +\lambda_{intra}\times L_{intra} \\
        L_{inter}&=Tanh(AR_{inter}+OFFSET_{inter}-AD_{inter})\\
        L_{intra}&=Tanh(AR_{intra}+OFFSET_{intra}-AD_{intra})
    \end{split}
\end{equation*}
where $L_{inter}$ and $L_{intra}$ are the atom distance losses for different elements and same elements in our 3 sets of base atom positions, respectively. Take first representation in Figure~\ref{fig:self-aug} as an example, we can find that the three sets are "Ca0-C1-O2", "Ca1-C0-O12" and "Ca2-C2-O20", respectively. $L_{inter}$ calculates the distance loss among three different elements for three sets. $L_{intra}$ calculates the distance loss among same elements. In this example, they are "Ca0-Ca1-Ca2", "C1-C0-C2" and "O2-O12-O20". By sampling the unique atom positions of three elements 32 times, we literally can calculate all the distances among atoms. The reason that we do not directly calculate all atom distance matrix is that symmetry operations on base atom sites involve uniqueness calculations, which is a time-intensive work and can slow down the model training process significantly. $\lambda_{inter}$ and $\lambda_{intra}$ are balancing parameters. $Tanh$ is hyperbolic tangent function. $AR_{inter}$ and $AR_{intra}$ are two atom radius sum. $OFFSET_{inter}$ and $OFFSET_{intra}$ are use to offset the atom radius sum. $AD_{inter}$ and $AD_{intra}$ are real distance. The goal of $L_{inter}$ and $L_{intra}$ is to push away two atoms and make the distance between two atoms at least bigger than $AD + OFFSET$.

We also set an upper bound for atom distances. The upper bound primarily limits the size of a unit cell and is defined as follow:

\begin{equation*}
    \begin{split}
        L_{box} = &\lambda_{volume}\times Cos(V_{r}, V_{f})\\ + &\lambda_{abc}\times (Cos(a_{r}, a_{f}) + Cos(b_{r}, b_{f}) + Cos(c_{r}, c_{f}))
    \end{split}
\end{equation*}
where $\lambda_{volume}$ and $\lambda_{abc}$ are balancing parameters. We calculate the cosine similarities between volume, unit cell lengths for real and fake materials in a minibatch.

\subsubsection{(d) Clustering and merging atoms of the same element to avoid colliding atoms}
\begin{figure}
	\centering
	\includegraphics[width=\linewidth]{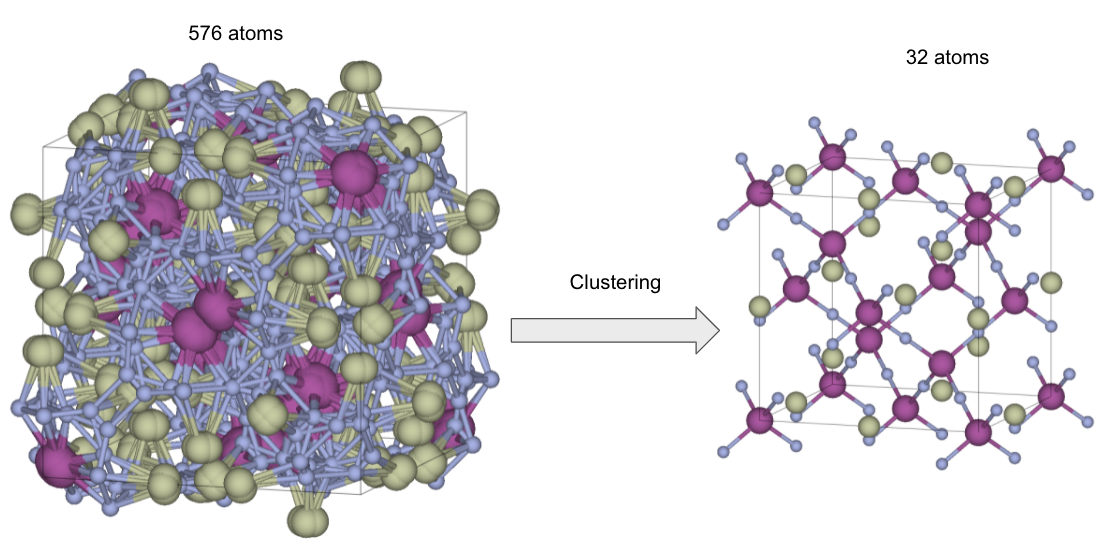}	
  	\caption {Clustering and merging overlapping atoms of the same elements}
  	\label{fig:cluster}
\end{figure}

One of the challenges for generating structures with high symmetry is that they tend to have a large number of symmetry operations. For example, Fd$\bar{3}$m has 192 symmetry operations, which easily generates large number of atom sites around a target site due to the slight coordinate deviation from the true site (Figure~\ref{fig:cluster} left). Here we propose a post-processing step to reduce the number of atoms in the unit cell. We cluster the nearby atoms of the same element and merge the elements that are crowd together using hierarchical clustering. Figure~\ref{fig:cluster} shows how the clustering and merging can reduce the number of atom in the unit cell with space group of Fd$\bar{3}$m from 576 to 32.

\subsection{DFT based validation}

% \hl{dilanga}
We apply the VASP \cite{Vasp1} based DFT calculation to relax each of the generated crystal structures and then calculate their formation energy to verify their dynamic stability. The details are the same as we did in \cite{song_computational_2021}.

\section{Results}

\subsection{Datasets}

We collect the training crystal structures from the Materials Project~\cite{MP_Database} and choose ternary materials with only 3 base atom sites. We use 12 space groups: 225 (4653 samples), 71 (2251), 221 (1467), 139 (970), 62 (900), 216 (811), 166 (784), 194 (753), 123 (485), 164 (470), 141 (435), 227 (421)).

\subsection{Training details}

Our losses consist of three parts: discriminator loss, generator loss and physics loss. Wasserstein distance is used to improve the model stability and balancing parameter for gradient penalty is 10.0 in this work. $\lambda_{inter}$ and $\lambda_{intra}$ are 3.5 and 5.0. $\lambda_{volume}$ and $\lambda_{abc}$ are 12.5 and 5.0. Learning rate for discriminator is 0.0002. Learning rate for generator and physics losses are both 0.0005. All training code are written using Pytorch.

\subsection{Generated crystal structures of different space groups}

\begin{figure*}
    \center
    \includegraphics[width=0.75\linewidth]{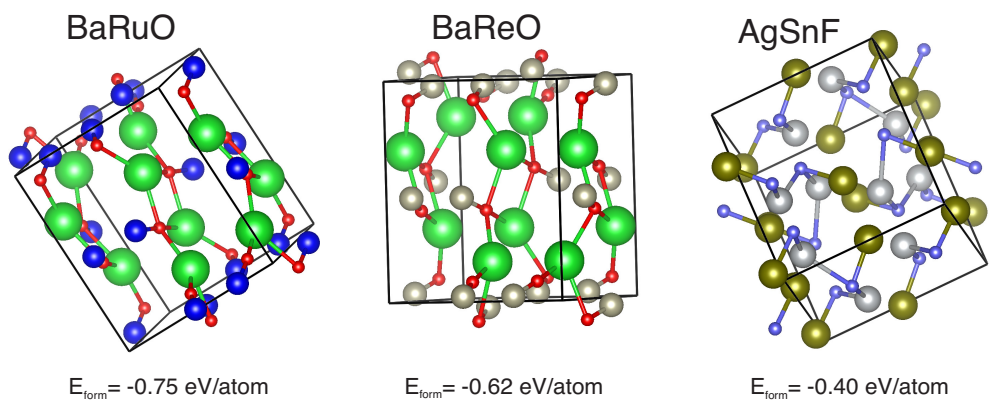}
    \caption{The structures and formation energies of BaRuO, BaReO and AgSnF orthorhombic materials generated by our GAN model.}
    \label{fig:gen-mat}
\end{figure*}

We trained generative model for 12 different space groups and generated 50,000 structures. Some of the generated stable crystal structures are shown in Figure~\ref{fig:gen-mat} along with their formation energy. 

\FloatBarrier
% \subsection{Effect of clustering for non-degenerative coordinates}

% \section{Discussion}

% cccc

\section{Conclusion} 

Deep learning based generative models have the potential to dramatically expand the current known materials repositories which can help to discover new materials with novel properties. While most generative models have been developed for computer vision or image/video/text generation, here we focus on the generation of atomic structures, the inorganic crystal materials. Here we propose a generic deep neural network based generative model for crystal structure prediction based on the Wasserstein generative adversarial network framework with physics oriented enhancements. Our study here shows that compared to pixels, the physical nature of atoms and their interactions make it necessary to introduce physics guided principles and constraints into the deep neural network generative models for generate more physically feasible materials. Our results show that these physical knowledge guided data augmentation, loss function design and post processing can greatly improve the performance of our generative models.

\section{ Acknowledgments}
Research reported in this work was supported in part by NSF under grants 1940099 and 1905775. The views, perspective, and content do not necessarily represent the official views of NSF.

\bibliographystyle{aaai}

\bibliography{references.bib}

% \bigskip
% \noindent Thank you for reading these instructions carefully. We look forward to receiving your electronic files!

\end{document}